\documentclass[twocolumn]{aastex62}
\usepackage{epsf}
\usepackage{natbib}
\usepackage{graphicx}
\usepackage{float}
\usepackage{afterpage}
\usepackage{amssymb}
\usepackage{enumitem}
\usepackage{bm,bbm,amssymb, amsmath}
\usepackage{url}
\usepackage[section]{placeins}

\usepackage{mathrsfs}
\usepackage{ulem}

\usepackage{hyperref}

\begin{document}
\title{\large Gravitational Capture of Large Interstellar Planetesimals by Collapsing Prestellar Cores}
\correspondingauthor{Richard Anderson}
\email{rbanderson@miners.utep.edu}

\author[0009-0004-7251-696X]{Richard Anderson}
\affil{University of Texas at El Paso, El Paso, TX, 79968, USA}

\author[0000-0003-1572-0505]{Kedron Silsbee}
\affil{University of Texas at El Paso, El Paso, TX, 79968, USA}

\begin{abstract}
We investigate the capture of free-floating planetesimals by the time-varying gravitational potential of a collapsing prestellar core.  We use one-dimensional collapse models to estimate the volume of space from which planetesimals are captured. We combine this calculation with an estimate of the size distribution and velocity dispersion of the free-floating planetesimal population to determine the largest planetesimal that is typically captured, as well as the total mass of all planetesimals captured by a given protostar. We consider two cases: one in which the free-floating planetesimal population comes from the birth cluster of the star, and one in which the planetesimals are captured from the Galactic population of free-floating planetesimals.  We find that the details of the collapse matter substantially, with the predicted diameter of the largest captured body from the Galactic population ranging from $2.8$ km to $6.4$ km depending on collapse model, and the largest body from our model birth cluster environment ranging from $70$ km to $1200$ km, for assumed velocity dispersions ranging from 0.7 to 8.5 km/s.  The total solid mass captured by this mechanism does not exceed 0.07 Earth masses in all cases.  We discuss the effect of these captured objects on the planet formation process, and argue that while the total solid mass captured is likely not significant, the largest captured bodies may grow to planet size via accretion of pebbles present in their host disks.  This process may accelerate planet formation, or enable it in disks that have difficulty in forming their own planetesimals. 
\end{abstract}
\section{Introduction}
The origin of planetesimals remains a major open question in the theory of planet formation.  They are believed to form from the coagulation of sub-micron sized dust particles present in the interstellar medium which are advected into the protoplanetary disk.  For small ($\leq 1$ mm) grain sizes, dust may grow by grains sticking together in mutual collisions.  However, as the grains grow to larger sizes, a combination of collisional fragmentation and rapid radial drift of dust within the disk likely inhibits further collisional growth \citep{Birnstiel12}.  Several effects have been proposed to concentrate dust grains in particular regions of the disk, thus accelerating the formation of planetesimals.  These include vortices \citep{Tanga96}, particle-gas instabilities \citep{Youdin05}, and the presence of ice-lines \citep{Ida16}.  However, all of these mechanisms have issues and uncertainties associated with them, and the origin of planetesimals remains uncertain.
\par
It was recently suggested \citep{Grishin19, Pfalzner19} that planetesimals ejected from disks surrounding other stars may be captured and act as seeds of planet formation, rather than planetesimals needing to form from scratch in each disk to initiate the planet-formation process.  So far, there have been detections of three such interstellar objects (which we term ``planetesimals" in the context of this paper), each a few hundred meters in size, passing through our solar system \citep{Meech17, Guzik20, Seligman25}.  In addition, two roughly meter-sized meteors of interstellar origin are reported to have impacted the Earth's atmosphere \citep{Siraj19, Siraj22}.  The abundance of such free-floating planetesimals is uncertain, but based on the detection of 'Oumuamua, \citet{Do18} estimated there to be $\sim 10^{15}$ such objects per cubic parsec in the local interstellar medium.  They estimate that this requires the ejection of approximately 40 earth masses of material per star, though this estimate can vary by orders of magnitude, depending on the assumed size distribution of such objects, or the exact size and density of 'Oumuamua.  This is not an unreasonable number: it is thought, based on modeling the formation of the Oort cloud \citep{Duncan87, Brasser10} in combination with estimates of its current mass \citep{Weissman83} that following the formation of the solar system, several tens of earth masses of solid material were ejected into interstellar space due to gravitational scattering by the giant planets.
\par
In order for a planetesimal to be captured in a disk around a star, it must achieve a negative total (kinetic plus potential) energy in the reference frame co-moving with the star.  This may occur either due to gas drag, or gravitational effects.  The role of gas drag has been explored in \citet{Grishin19} and \citet{MoroMartin22}.  \citet{MoroMartin22} find that planetesimals up to 10 km in diameter can be captured by this mechanism, but that it is inefficient for larger bodies.
\par
The second way a planetesimal can lose energy is due to the time variability of the gravitational potential well associated with the forming star.  As the prestellar core collapses to form a protostar, the gravitational potential in the vicinity of the star becomes more negative.  This means that a planetesimal that is crossing the region during the collapse must climb out of a deeper potential well than it fell into, and thus will lose total energy.  This is analogous to the Integrated Sachs-Wolfe effect on the cosmic microwave background \citep{Sachs67}, in which photons crossing clusters of galaxies may be red or blue shifted, depending on how the cluster potential is evolving.  This effect, unlike that of gas drag, is independent of the size of the planetesimal, and therefore presents the possibility of capturing planetesimals hundreds or thousands of km in diameter. 
\par
The capture of large planetesimals is of particular interest because even one large planetesimal may continue to grow to planetary size, via rapid gas-drag assisted accretion of centimeter-sized dust grains in the disk known as pebble accretion.  The minimum diameter required for a planetesimal to grow to planet size within a typical disk lifetime of ~$10^6$ years is thought to be on the order of a few hundred to a thousand kilometers \citep{Lambrechts12, Lyra23}, depending on the assumed parameters of the disk and pebble populations. Capture of planetesimals due to the time-varying gravitational potential is the subject of the present paper.
\par
\citet{Pfalzner21} studied gravitational capture via simulations of gas and planetesimal dynamics in a collapsing and fragmenting molecular cloud.  They found that a highly variable fraction - over 10\% in one of their simulations, but less than $10^{-4}$ in others, of the planetesimals become gravitationally bound to individual overdensities, which are expected to collapse and form stars.  However, in that study, the planetesimals were initially at rest relative to the molecular cloud, so it remains an open question what effect a realistic initial velocity dispersion would have on the capture rate of planetesimals.
\par
In the current paper, we analyze three one-dimensional models for the collapse of a prestellar core, and use analytic methods and numerical simulations to analyze the planetesimal capture rate in each case.  The paper is organized as follows.  In Section \ref{sect:homoCollapse} we analyze the collapse of a homogeneous cloud of gas at a fixed fraction of the free-fall speed.  We derive the scaling of the capture rate on initial velocity of the planetesimal, and confirm our results with numerical simulations.  In Section \ref{sect:sss}, we analyze the capture of planetesimals in the self-similar models of \citet{Larson69} and \citet{Shu77}.  In Section \ref{sect:biggestCapture} we combine the capture rate results from Sections \ref{sect:homoCollapse} and \ref{sect:sss} with models for the velocity dispersion and size distribution of planetesimals to determine the largest planetesimal typically captured by a forming proto-star.  We consider both the Galactic population of free-floating planetesimals, as well as a denser, slower-moving population of planetesimals from the birth-cluster of the protostar.    We present our conclusions in Section \ref{sect:conclusion}.

\section{Homogeneous Collapse Model}
\label{sect:homoCollapse}
In this section we assume a homogenous density profile which is collapsing at a fixed fraction of the free-fall rate.  In particular, we assume the radius of the cloud to satisfy 
 \begin{equation}
 \dot R = -\alpha \sqrt{\frac{2GM}{R}},
 \label{eq:collapseVelocity}
 \end{equation}
  with $0 \leq \alpha \leq 1$.  We define time $t = 0$ to be the point at which $R = 0$, and therefore the collapse process corresponds to times $t < 0$.
  \par
Observations of prestellar cores in the initial stages of collapse find that their lifetime at a particular density is a few times the free-fall timescale at that density, with this ratio generally decreasing towards higher density in the range of $10^4 - 10^6$ particles per cm$^3$; see Figure 2 of \citet{WardThompson07} or Figure 9 of \citet{Konyves15}. While this is not the most realistic model of the collapse of a prestellar core, it admits a semi-analytic expression for the rate at which planetesimals are captured, and serves as a good test of our numerical procedures.
\par
 To determine the rate at which the potential changes, we first integrate Equation \eqref{eq:collapseVelocity} to obtain $R$ as a function of time:
 \begin{equation}
 R = \left[\frac{-3 \alpha \sqrt{GM} t}{\sqrt{2}}\right]^{2/3}.
 \label{eq:R}
 \end{equation}
 We then calculate the potential as a function of coordinate $r$ and cloud radius $R$ as
 \begin{equation}
 \phi(r, R) = 
 \begin{cases} 
      \frac{-GM}{2R}\left[3-\left (\frac{r}{R}\right)^2\right] & 0 < r < R \\
      -\frac{GM}{r} & r > R.
   \end{cases}
   \label{eq:phi}
\end{equation}

As we are ignoring any non-gravitational forces on the planetesimal, its kinetic energy satisfies
\begin{equation}
\dot E_k = -\vec v \cdot \nabla \phi,
\label{eq:kineticEnergy}
\end{equation}
where $v$ is the particle's velocity, and the specific potential energy satisfies
\begin{equation}
\dot E_p = \vec v \cdot \nabla \phi + \frac{\partial \phi}{\partial t}.
\label{eq:potentialEnergy}
\end{equation}
Adding Equations \eqref{eq:kineticEnergy} and \eqref{eq:potentialEnergy} gives
\begin{equation}
\frac{dE_{\rm tot}}{dt} = \frac{\partial \phi}{\partial t}.
\end{equation}
We can calculate 
\begin{equation}
\frac{\partial \phi}{\partial t} = 
\begin{cases}
\frac{\partial \phi}{\partial R} \frac{\partial R}{\partial t} = \frac{-3 \alpha}{\sqrt{2}} \sqrt{\frac{(GM)^3}{R^5}} \left[1 - \left(\frac{r}{R}\right)^2 \right] & 0 < r < R\\
0 & r \geq R.
\end{cases}
\end{equation}
\par
Let us then consider what happens to particles that pass through the collapsing cloud.  For analytic estimates, we assume the impact parameter to be 0 (i.e. particles go straight through the center of the cloud).  We later relax this assumption when performing numerical simulations. We define $v_0$ to be the initial speed of the approaching particle when it is infinitely far from the collapsing cloud.  We furthermore define $R_*$ as the value of $R$ at the time the particle reaches the edge of the cloud and $v_* = \sqrt{GM/R_*}$.   We then consider two limits: particles for which $v_0 \gg v_*$, and those for which $v_0 \ll v_*$.  
\subsection{Rapid Initial Approach: $v_0 \gg v_*$}
For the case in which $v_0 \gg v_*$, we approximate the trajectory of the particle as unperturbed by the potential, and that $R$ is equal to $R_*$ for the time that the particle crosses the cloud.  In this case, the change in the particle's specific energy as it traverses the cloud is approximated by
\begin{equation}
\begin{split}
\Delta E &= \frac{2}{v_0} \int_0^{R_*} \frac{-3 \alpha}{\sqrt{2}} \sqrt{\frac{(GM)^3}{R_*^5}} \left[1 - \left(\frac{r}{R_*}\right)^2 \right]dr\\
&= \frac{-2\sqrt{2}\alpha v_*^3}{v_0}
\end{split}
\end{equation}
Normalizing this by the initial specific energy $E_i = v_0^2/2$ gives 
\begin{equation}
\Delta \tilde E \equiv \frac{|\Delta E|}{E_i} = 4 \sqrt{2} \alpha \left(\frac{v_*}{v_0}\right)^3.
\end{equation}
Since we assumed $v_0 \gg v_*$, then the energy lost is less than the initial energy of the particle, so it is not captured.

\subsection{Slow Initial Approach: $v_0 \ll v_*$}
Particles with $v_0 \ll v_*$, are strongly perturbed by the gravitational potential of the collapsing cloud.  Furthermore, the time that the particle spends within the cloud depends somewhat on $\alpha$.  We consider here the case in which $\alpha \ll 1$, and therefore the value of $R$ does not change significantly during the time it takes the particle to cross the cloud.  In this case, we estimate the lost specific energy (again normalized to the initial specific kinetic energy) as 
\begin{equation}
\begin{split}
\Delta \tilde E &= \frac{4}{v_0^2} \int_0^{R_*} v(r)^{-1} \left|\frac{\partial}{\partial t} \phi(r, R_*)\right| dr \\ &= \frac{4}{v_0^2} \int_0^{R_*} \frac{3\alpha}{\sqrt{2}} \left(\frac{GM}{R_*}\left[3-\left(\frac{r}{R_*}\right)^2\right]\right)^{-1/2} \\
& \times \sqrt{\frac{(GM)^3}{R_*^5}} \left[1-\left(\frac{r}{R_*}\right)^2\right]dr \\
&= 6 \sqrt{2} \alpha \frac{v_*^2}{v_0^2} \int_0^1 \frac{1-x^2}{\sqrt{3-x^2}} dx\\
 &\approx 3.4 \alpha \left(\frac{v_*}{v_0}\right)^2.
 \end{split}
\label{eq:slowEnergyLoss}
\end{equation}
Let us assume now that during its collapse, the cloud is embedded in a sea of planetesimals with velocity $v_0$ and number density $n_0$.  Particles are captured if $\Delta \tilde E > 1$. Equation \eqref{eq:slowEnergyLoss} shows this criterion to be equivalent to the condition that $R_* < R_{\rm max}(v_0)$, where
\begin{equation}
    R_{\rm max}(v_0) = 3.4 \alpha GM/v_0^2.
\end{equation}
The encounter rate $\mathscr{R}$, including gravitational focusing is given as a function of cloud radius $R$ by 
\begin{equation}
\mathscr{R}(R) = \pi R^2 n_0 v_0 \left[1+ \left(\frac{2v_*}{v_0}\right)^2\right] \approx \frac{2\pi n_0 R GM}{v_0}.
\end{equation}
The total number captured can then be written as a function of $v_0$ as
\begin{equation}
N_{\rm capt}(v_0) = \int_0^{R_{\rm max}} \frac{\mathscr{R}(R)}{|\dot R|} dR = 38 \alpha^{3/2} n_0 \left(\frac{GM}{v_0^2}\right)^3.
\label{eq:Ncaptv0}
\end{equation}

\subsection{Average over Maxwellian Distribution for $v_0$}  
If we posit a specific distribution for $v_0$, then we can use Equation \eqref{eq:Ncaptv0} to determine the number of captured particles:
\begin{equation}
N_{\rm capt}^{\rm tot} = \int_0^\infty N_{\rm capt}(v_0) f(v_0) dv_0.
\label{eq:totalCaptured1}
\end{equation}
We assume the planetesimals to have a Maxwellian speed distribution with respect to the collapsing prestellar cores, i.e.
\begin{equation}
    f(v_0;\sigma) = \sqrt{\frac{2}{\pi}}\,\frac{v_0^{2}}{\sigma^{3}}
    \exp\left(-\frac{v_0^{2}}{2\sigma^{2}}\right),
    \label{eq:Maxwellian}
\end{equation}
where $\sigma$ is the one-dimensional velocity dispersion.
If we substitute Equations \eqref{eq:Ncaptv0} and \eqref{eq:Maxwellian} into Equation \eqref{eq:totalCaptured1}, we find that the integral diverges at the lower bound.  To address this, we introduce an initial value of $R$, termed $R_{\rm init}$, from which the gravitational collapse is taken to start.  We then replace the upper bound of the integral in Equation \eqref{eq:Ncaptv0} with ${\rm min}(R_{\rm max}, R_{\rm init})$ to get
\begin{equation}
\begin{split}
N_{\rm capt}(v_0) &= \int_0^{{\rm min}(R_{\rm max}, R_{\rm init})} \frac{\mathscr{R}}{|\dot R|} dR \\
\\
&=
 \begin{cases} 
     38 \alpha^{3/2} n_0 \left(\frac{GM}{v_0^2}\right)^3& v_0 > v_{\rm crit} \\
    \frac{4\pi n_0 \sqrt{GM}}{5\sqrt{2} \alpha v_0} R_{\rm init}^{5/2} & v_0 < v_{\rm crit},
   \end{cases}
   \end{split}
\end{equation}
where $v_{\rm crit} = \sqrt{3.4 \alpha GM/R_{\rm init}}$ is the critical velocity which is marginally captured if it impacts the cloud when it has size $R_{\rm init}$.
With this assumption, Equation \eqref{eq:totalCaptured1} gives a finite answer:
\begin{align}
& N_{\rm capt}^{\rm tot} = \int_0^\infty N_{\rm capt}(v_0) f(v_0) dv_0 = 
\nonumber \\
 & \int_0^{v_{\rm crit}} \frac{4\pi n_0 \sqrt{GM}}{5\sqrt{2} \alpha v_0} R_{\rm init}^{5/2}  \left[\frac{4 \pi v_0^2}{\left(\sqrt{2\pi}\sigma\right)^3} e^{\frac{-v_0^2}{2\sigma^2}}\right] dv_0 + \nonumber \\ 
 &\int_{v_{\rm crit}}^\infty \frac{4 \pi n_0 v_0^2 e^{\frac{-v_0^2}{2\sigma^2}}}{\left(\sqrt{2\pi}\sigma \right)^3} \cdot 38 \alpha^{3/2} \left(\frac{GM}{v_0^2}\right)^3dv_0
\nonumber \\
& = 0.71 \frac{n_0}{\alpha} R_{\rm init}^3 \frac{v_{\rm crit}^2 v_*^{\rm init}}{\sigma^3} + 10 \alpha^{3/2} n_0 \left(\frac{GM}{v_{\rm crit} \sigma}\right)^3,
\label{eq:totalCaptured3}
\end{align}
where $v_*^{\rm init} = \sqrt{GM/R_{\rm init}}$.  In evaluating the integrals in Equation \eqref{eq:totalCaptured3}, we have assumed $\sigma \gg v_{\rm crit}$, so that the exponential factors may be replaced with unity. 
\par
Using the definitions of $v_{\rm crit}$ and $R_{\rm init}$, we see that 
\begin{equation}
\frac{GM}{v_{\rm crit}^2}  = \frac{0.30}{\alpha}R_{\rm init}.
\end{equation}
Furthermore, 
\begin{equation}
\frac{v_*^{\rm init}}{v_{\rm crit}} = .54/\sqrt{\alpha}
\end{equation}
We can therefore re-write the final line of Equation \eqref{eq:totalCaptured3} as 
\begin{equation}
N_{\rm capt}^{\rm tot} = 0.64 \frac{n_0 R_{\rm init}^3}{\alpha^{3/2}} \left(\frac{v_{\rm crit}}{\sigma}\right)^3.
\label{eq:finalAnswer}
\end{equation}
  Plugging in $v_{\rm crit} = \sqrt{3.4 \alpha GM/R_{\rm init}}$ lets us write 
\begin{equation}
N_{\rm capt}^{\rm tot} = 4.0 n_0 \left(\frac{\sqrt{GMR_{\rm init}}}{\sigma}\right)^3.
\end{equation}
Interestingly, this is independent of the value of $\alpha$ (subject to our assumption that $\alpha \ll 1$).

\subsection{Numerical Verification of Results}
\label{sect:homoSim}
We ran simulations of planetesimal trajectories using the REBOUND code \citep{Rein12} with the IAS15 integrator \citep{Rein15}, taking $M=M_\odot$ and $\alpha=0.2$.  We ran a suite of  simulations of planetesimals approaching and passing through a collapsing cloud.  Simulations varied in the initial velocity and range of possible impact parameters of the incoming planetesimals.  Each simulation proceeded in two phases.  In phase 1, lasting for time $2R_{\rm init}/\sqrt{v_0^2+2GM/R_{\rm init}}$ (an upper limit on the time for a particle to cross the cloud), the gravitational potential was held fixed, so that the cloud would be populated with a realistic distribution of initial particle positions and velocities.  During this initial phase, we allowed particles with impact parameters between $b_1$ and $b_2$ to enter the cloud at a constant rate given by 
\begin{equation}
\dot N_{\rm static} =  \pi \left(b_2^2 - b_1^2 \right)v_0 n_0.
\end{equation}
After this initial phase, we allowed the gravitational potential to evolve as described in Equations \eqref{eq:R} and \eqref{eq:phi}. At this point, the rate at which particles enter the cloud is modified to account for the recession of the cloud surface away from the incoming particle flux:
\begin{equation}
    \dot N = \dot N_{\rm static} \max \left[(1 - \dot R/v_r), 0 \right],
    \label{eq:actual_entry_rate}
\end{equation}
where 
\begin{equation}
   v_r = \sqrt{\frac{2GM}{R} + v_0^2\left(1-\frac{b^2}{R^2}\right)} 
\end{equation}
is the radial velocity of the particle at the cloud surface.
We define the capture volume as $V_{\rm capt} = N_{\rm captured}/n_0$.  This is the total volume of space from which planetesimals are captured over the entire collapse process.  Note that we obtain the same expected capture volume independent of the value of $n_0$ that we assume in Equation \eqref{eq:actual_entry_rate}. 
\par
We ran simulations for 79 ranges of $b_i$ evenly separated in log-space between $10^9$ and $10^{17}$ meters.  The upper limit of this range is larger than the maximum value of $b$ for particles that would impact the cloud, and therefore does not affect our results.  We do miss a small number of particles by not considering impact parameters less than $b_{\rm min} = 10^9$ meters, however even if every particle that entered the cloud with $b < 10^9$ meters were captured, this would have a negligible impact on the capture volume.
\begin{figure}[!htbp]
    \centering
\includegraphics[width=\columnwidth]{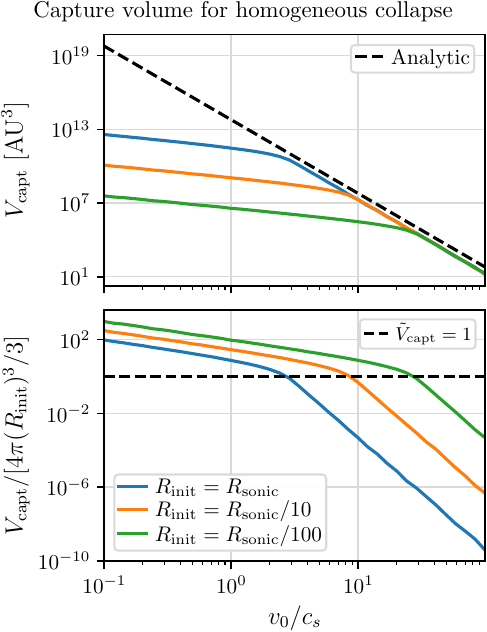}
    \caption{Top panel: volume $V_{\rm capt}$ of interstellar space from which planetesimals are captured as a function of their initial velocity with respect to the cloud normalized by the sound speed. We assume $\alpha=0.2$, $M = 1 M_\odot$ for all curves.  The initial radius is taken to be either $R_{\rm sonic}$, $R_{\rm sonic}/10$ or $R_{\rm sonic}/100$, where $R_{\rm sonic}$ = 2,110 AU is the radius at which $\dot R = -c_s$.  Different color curves correspond to different initial cloud radii $R_{\rm init}$. The analytic approximation given in Equation \eqref{eq:Ncaptv0} is shown by the dashed black curve. Bottom panel: $V_{\rm capt}$ normalized by the initial cloud volume $4\pi R_{\rm init}^3/3$. The horizontal dashed black line indicates a normalized capture volume of 1.}
    \label{fig:combinedCaptureVolume}
\end{figure}
\par
The top panel of Figure \ref{fig:combinedCaptureVolume} shows the capture volume as a function of the initial velocity $v_0$ of the planetesimal.  In addition, we have plotted the analytic approximation for the capture volume given by taking $N_{\rm capt}$ from Equation \eqref{eq:Ncaptv0}, and dividing by $n_0$.  Different curves correspond to different values of the initial radius, expressed in terms of $R_{\rm sonic}$, where $R_{\rm sonic} = 2 \alpha^2 GM/c_s^2$ = 2,110 AU is the radius $R$ at which $\dot{R} = -c_s$.  The sound speed is taken to be 183 m/s, corresponding to a temperature of approximately 9.5 Kelvin, and the cloud mass $M$ is taken to be one solar mass.    
\par
The bottom panel shows the capture volumes normalized by the initial volume of the cloud $V_{\rm init} = 4 \pi R_{\rm init}^3/3$.  As can be seen, the cloud captures its own volume roughly when $v_0$ is equal to the escape speed of the cloud.  In this panel, the dashed line is just a horizontal line at $V_{\rm capt}/V_{\rm init} = 1$, drawn to guide the eye.

\section{self-similar solutions}
\label{sect:sss}
While the process of cloud-collapse is complicated and there is not an agreed-upon best analytic model, there are two widely-discussed 1D self-similar solutions in the literature; the Larson-Penston solution \citep{Larson69, Penston69}, and the Shu inside-out collapse model \citep{Shu77}.  In this section, we numerically calculate the rate of planetesimal capture, assuming the gas density distribution to evolve according to each of these collapse models.  In these cases, we restrict ourselves to a numerical calculation of the capture rate, however we do provide an analytic estimate of the maximum value of $v_0$ that still permits capture.

\subsection{Larson-Penston Solution}
\label{sect:Larson}
In this model, the density is given by
\begin{equation}
\rho(r, t) = \frac{\eta\left(\frac{-r}{c_st}\right)}{4\pi Gt^2},
\label{eq:rhoLarson}
\end{equation}
for $t \leq 0$.  $\eta$ is given approximately by 
\begin{equation}
 \begin{cases} 
 \eta(x) = \frac{B}{A^2} & x < A \\
 \eta(x) = B/x^2 & x > A,
 \end{cases}
 \label{eq:etaAnalytic}
 \end{equation}
Where A = 2.31, and B = 8.86.  Because the density only falls off as $r^{-2}$, this solution formally has infinite mass.  We truncate the solution at the radius $R_{\rm max}$ such that the enclosed mass is equal to one solar mass.  That is
\begin{equation}
    \int_{r = 0}^{R_{\rm max}} 4 \pi r^2 \rho(r) dr = 1 M_\odot.
    \label{eq:integralMassConstraint}
\end{equation}

\subsubsection{Behavior as $t \rightarrow 0$}
\label{sect:LarsonLimit}
Intuitively one might suspect, as in the case of a homogeneous collapse, that arbitrarily fast-moving planetesimals may be captured because the potential changes very rapidly near $t = 0$, however this is not the case.  The potential at the center of a spherically symmetric mass distribution is given by 
\begin{equation}
    \phi(0) = -\int_0^{R_{\rm max}} \frac{4 \pi r^2 G \rho(r) dr}{r}.
    \label{eq:phiLarson1}
\end{equation}
Using Equation \eqref{eq:rhoLarson}, we can re-write Equation \eqref{eq:phiLarson1} as 
\begin{equation}
    \phi(0) = -c_s^2 \int_0^{x_{\rm max}} x \eta(x) dx,
\end{equation}
where $x = -r/(c_s t)$.  We can then write $\dot \phi(0)$ as 
\begin{equation}
    \dot \phi(0) = -c_s^2 x_{\rm max} \eta(x_{\rm max}) \dot x_{\rm max},
    \label{eq:dotphiLarson}
\end{equation}
where
\begin{equation}
\dot x_{\rm max} = \frac{-\dot R_{\rm max}}{c_s t} + \frac{R_{\rm max}}{c_st^2} = \frac{-\dot R_{\rm max}}{c_s t} - \frac{x_{\rm max}}{t}. 
\end{equation}
Since we are analyzing the behavior as $t \rightarrow 0$, we drop the first term in favor of the second.  As $t \rightarrow 0$, $x_{\rm max} \rightarrow \infty$, and therefore $\eta(x_{\rm max})\rightarrow B/x_{\rm max}^2$.  In this limit, Equation \eqref{eq:dotphiLarson} becomes
\begin{equation}
\dot \phi(0) = \frac{B c_s^2}{t}.
\end{equation}
We then ask how much energy is lost in crossing the central region of the cloud in which $x < A$.  The crossing time is given by 
\begin{equation}
    \tau_{\rm cross} = \frac{-2Ac_st}{\sqrt{-2 \phi(0)}} \approx \frac{-2 A t}{ \sqrt{2B\ln{x_{\rm  max}}}}.
\end{equation}

Therefore, the amount of specific energy lost by the particle in crossing the sphere of radius $-Ac_s t$ is approximately
\begin{equation}
    \Delta E \approx \dot \phi(0) \tau_{\rm cross} = \frac{-2ABc_s^2}{\sqrt{2B\ln(x_{\rm max})}}.
\end{equation}
The specific energy lost in the central region of the cloud remains of order $c_s^2$, and actually approaches zero as $t = 0$, since the denominator diverges logarithmically.
\par
Since the critical velocity for capture is not determined by the limiting behavior of the potential near t = 0, as a very rough estimate of the critical velocity for capture, we estimate the change of specific energy as 
\begin{equation}
\Delta E = \tau_{\rm cross}\dot \phi(R_{\rm max}).
\end{equation}
Here $\dot \phi(R_{\rm max}) = GM\dot R_{\rm max}/R_{\rm max}^2$, and we can approximate
\begin{equation}
    \tau_{\rm cross} = \frac{2R_{\rm max}}{\sqrt{v_{\rm esc}^2 + v_0^2}}.
\end{equation}
In the limit that $x_{\rm max} \gg 1$, Equations \eqref{eq:rhoLarson} and \eqref{eq:etaAnalytic} show that $v_{\rm esc} = \sqrt{2B} c_s.$
  Conservation of mass requires that $\dot R_{\rm max}$ satisfy the equation 
\begin{equation}
    4 \pi R_{\rm max}^2 \rho(R_{\rm max}) \dot R_{\rm max} = -\int_0^{R_{\rm max}} 4 \pi r^2 \dot \rho(r) dr.
    \label{eq:rdotEquation}
\end{equation}
Solving Equation \eqref{eq:rdotEquation} using the analytic approximation in Equation \eqref{eq:etaAnalytic} would suggest $\dot R_{\rm max} = - 2c_s A/3 = -1.54 c_s$, however a more accurate calculation using the numerical solution for $\eta$ yields $\dot R_{\rm max} = -\psi c_s$, with $\psi = 3.28$.
\par
The maximum value of $v_0$ that can be captured is obtained by letting $\Delta E = -v_{\rm max}^2/2$.  Letting $v_{\rm max} = \chi c_s$, this relation reduces to the equation 
\begin{equation}
    \chi^2 = \frac{4 B \psi}{\sqrt{2B + \chi^2}} \rightarrow \chi = 4.4.
    \label{eq:LarsonCriticalVelocity}
\end{equation}

\subsubsection{Numerical Results}
We ran a set of simulations using the algorithm described in Section \ref{sect:homoSim}, but assuming the gas density within the cloud to be described by Equation \eqref{eq:rhoLarson} for $0 < r < R_{\rm max}$ and 0 for $r > R_{\rm max}$.  We did not use the approximate expressions for $\eta$ given in Equation \eqref{eq:etaAnalytic}, but instead the numerical solution of Equations C6 from \citet{Larson69} subject to the boundary conditions described in that paper.  There are two time-dependent characteristic radii in this model: $R_{\rm max}$, defined by Equation \eqref{eq:integralMassConstraint} and $R_0 = -2.31c_st$, which approximates the radius at which the density profile transitions from being flat to declining as $1/r^2$.  We define $R_{\rm eq} = 1.34 \times 10^4$ AU to be the radius at which $R_0 = R_{\rm max}$. We ran three separate simulations beginning when $R_0 = R_{\rm eq}$, $R_0 = R_{\rm eq}/10$ and $R_0 = R_{\rm eq}/100$.
\par
\begin{figure}[!htbp]
    \centering
    \includegraphics[width=\columnwidth]{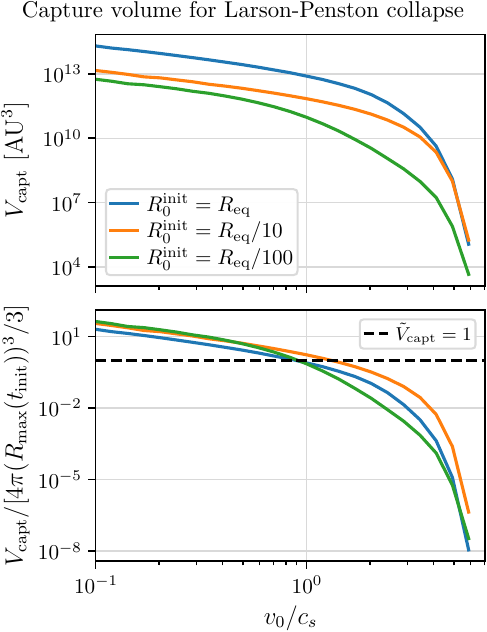}
    \caption{Top panel: volume $V_{\rm capt}$ of interstellar space from which planetesimals will be captured as a function of their initial velocity with respect to the cloud, normalized by the sound speed. This assumes that the potential evolves as in the Larson-Penston solution described in Section \ref{sect:Larson}. Different color curves correspond to initial scale radii $R_0^{\rm init}=R_{\rm eq}$, $R_{\rm eq}/10$, and $R_{\rm eq}/100$; see text for details. Bottom panel: $V_{\rm capt}$ normalized by the volume of the truncated cloud $4 \pi R_{\rm max}^3(t_{\rm init})/3$. The horizontal dashed black line indicates a normalized capture volume of 1.}
    \label{fig:LarsonCaptureRate}
\end{figure}

  The results are shown in Figure \ref{fig:LarsonCaptureRate}.  Different color lines correspond to different times at which the simulation was begun, characterized by the value of $R_0$ at that time.
 The capture rate drops to zero for values of $v_0$ greater than $\sim 6 c_s$, in rough agreement with the value of 4.5 predicted by Equation \eqref{eq:LarsonCriticalVelocity}.  As in the homogeneous case, the capture volume increases quickly for smaller values of $v_0$.  The bottom panel shows the capture volume normalized by the volume of the cloud core, $4 \pi R_{\rm max}^3/3$, evaluated at the initial time.  As in the homogeneous case, the capture volume is roughly equal to the cloud volume for planetesimals with $v_0 \approx c_s$.

\subsection{Shu Solution}
\label{sect:Shu}
The final stage of star formation is the accretion of material onto the young protostar.  We model that using the self-similar ``expansion-wave collapse solution" presented in \citet{Shu77}.  In this case, there is a central point mass that grows with time, surrounded by a continuous spherically symmetric envelope.  This envelope has density given by
\begin{equation}
\rho(r, t) = \frac{\alpha(x)}{4\pi Gt^2},
\label{eq:rhoShu}
\end{equation}
and the infall velocity is given by 
\begin{equation}
v(r, t) = c_s \tilde v(x),
\label{eq:vShu}
\end{equation}
where $x = r/(c_s t)$.  As in our other models, we assume the total cloud mass to be 1 $M_\odot$.  This implies a maximum radius $R_{\rm max} = GM_\odot/(2c_s^2) = 1.32\times10^4$ AU.  The functions $\alpha$ and $\tilde v$ are given by numerical solution of Equations 11 and 12 from \citet{Shu77} [note that what we call $\tilde v$ here is negative 1 times $v$ from \citet{Shu77}]. $\alpha(x)$ has the property that for $x \ll 1$, $\alpha(x) = 0.70 x^{-3/2}$, and for $x > 1$, $\alpha(x) = \frac{2}{x^2}$.  For $x \ll 1$, $\tilde v(x) = 1.4/\sqrt{x}$ and for $x > 1$, $\tilde v(x) = 0$.  This solution is defined for $t > 0$, and for $t < 0$, the density profile is inversely proportional to $r^2$.
\par
While the solution from \citet{Larson69} approaches a $1/r^2$ density profile as $t \rightarrow 0$, and the solution from \citet{Shu77} begins with a $1/r^2$ density profile, the end state of the Larson-Penston solution does not perfectly match the initial condition for the Shu solution, since the normalization constant differs by a factor of 4.4.  For that reason, in this work, we treat these as two separate models, rather than as two stages of the same process.
\par
The accretion rate onto the point mass at the center of the Shu solution is given by 
\begin{equation}
\dot M(t) = \lim_{\rm r \rightarrow 0} 4\pi r^2 \rho(r, t) v(r, t).
\label{eq:MdotLimit}
\end{equation}
Using Equations \eqref{eq:rhoShu}, \eqref{eq:vShu}, and \eqref{eq:MdotLimit} plus the limiting forms for $\alpha$ and $\tilde v$, we find that 
\begin{equation}
\dot M(t) = \frac{0.98c_s^3}{G} \rightarrow M(t) = \frac{0.98c_s^3t}{G} = \frac{0.98 c_s^2 R_0}{G}, 
\label{eq:centralMass}
\end{equation}
where $R_0$ is now defined by $R_0 = c_st$.  Using Equation \eqref{eq:rhoShu}, in the limit that $t = 0$, we find that the {\it initial} mass present within radius $R_0$ is equal to $2 c_s^2 R_0/G$.  Comparison with Equation \eqref{eq:centralMass} shows that at all times, 49\% of the mass within radius $R_0$ is contained in the central body.  
\subsubsection{Behavior as $t \rightarrow 0$}
\label{sect:ShuLimit}
We perform an analysis for the Shu collapse model similar to that done in Section \ref{sect:LarsonLimit} for the Larson-Penston model.  We first note that the potential is static outside of $R_0$.  To estimate the potential interior to $R_0$, we approximate that all material interior to $R_0$ is contained in the central point-mass. As seen in Equation \eqref{eq:centralMass} this overestimates the central mass by a factor of 2.0, thus overestimating the potential by at most a factor of 2.  We add the potential from the point-mass to that from the material outside of $R_0$.  This gives an approximate potential 
\begin{equation}
    \phi(r) = -2c_s^2 \left[\frac{R_0}{r} + \chi \right],
    \label{eq:ShuPot}
\end{equation}
where $\chi = \ln{\left(\frac{R_{\rm max}}{R_0}\right)}$
Remembering that $\dot R_0 = c_s$, we can calculate 
\begin{equation}
    \dot \phi = 2c_s^3 \left[\frac{1}{R_0} - \frac{1}{r}\right].
    \label{eq:ShuPhidot}
\end{equation}
We then note that 
\begin{align}
    \Delta E &= 2 \int_0^{R_0} \frac{\dot \phi(r)}{\sqrt{-2 \phi(r)}}dr \nonumber\\ &= 2 c_s^2 \left[\frac{\sqrt{1+\chi}}{\chi} - \frac{1+2\chi}{\chi^{3/2}} \sinh^{-1} \sqrt{\chi}\right].
\end{align}
Similar to the Larson case, this function goes to zero as $t \rightarrow 0$ (since $\chi \rightarrow \infty$ in this limit).
\par
To estimate the critical capture velocity, we attempt a similar approach as for the Larson-Penston solution, and estimate  the energy lost when the particle crosses the entire cloud.  The natural estimate would be 
\begin{equation}
    \Delta E = \tau_{\rm cross}( R_0) \dot \phi (R_0). 
\end{equation}
However, since $\dot \phi(R_0) = 0$ [see Equation \eqref{eq:ShuPhidot}], we modify this approach slightly, and instead calculate 
\begin{equation}
        \Delta E(\delta) = \tau_{\rm cross}(\delta R_0) \dot \phi (\delta R_0). 
        \label{eq:DeltaEShu}
\end{equation} 
and set this equal to $-v_0^2/2$, and solve for $v_0$.  We then choose the value of $\delta$ that maximizes the magnitude of $\Delta E$.
Using Equations \eqref{eq:ShuPot} and \eqref{eq:ShuPhidot}, Equation \eqref{eq:DeltaEShu} yields 
\begin{equation}
    \frac{v_0^2}{2} = \frac{4(1 - \delta) c_s^2}{\sqrt{\frac{v_0^2}{c_s^2} + 4 \left(\frac{1}{\delta} + \chi\right)}}
\end{equation}
Substituting $s = (v_0/c_s)^2$, this simplifies to 
\begin{equation}
s^3 + 4s^2\left(\chi + \frac{1}{\delta}\right) - 64 \left(1 - \delta \right)^2 = 0
\label{eq:ShuVCrit}
\end{equation}
For a given value of $\delta$, this is maximized for the smallest allowable value of $\chi$.  Using $\chi = 1$, we solve Equation \eqref{eq:ShuVCrit} numerically, and find that the maximum value of s is 1.3, corresponding to $\delta = 0.27$.  From this analysis, we could estimate that objects with $v_0 > 1.1 c_s$ would not be captured in this model.  As seen in the following section, numerical simulations give a critical value of $v_0 = 1.4 c_s$.

\subsubsection{Simulations of the Shu Collapse Model}

We ran a set of simulations using REBOUND and the IAS15 integrator as described in Section \ref{sect:homoSim}, but using the mass distribution and potential of the Shu solution, including the growing central point mass.  Since the potential is static outside the expansion wave, particles enter the evolving region at $R_0=c_st$, with their initial velocities corrected for gravitational focusing and their entry rate augmented by the expansion of this boundary.  We followed the evolution from $t=10^{-9}t_{\rm final}$ to $t_{\rm final}=3.5\times10^5$, $3.5\times10^4$, and $3.5\times10^3$ years.  These times correspond to the expansion wave reaching approximately all, one tenth, and one hundredth of the initial cloud mass, respectively. We averaged over an expected number of 400 planetesimals injected per pair of initial velocity and impact parameter. 
\par
\begin{figure}[!htbp]
    \centering
\includegraphics[width=\columnwidth]{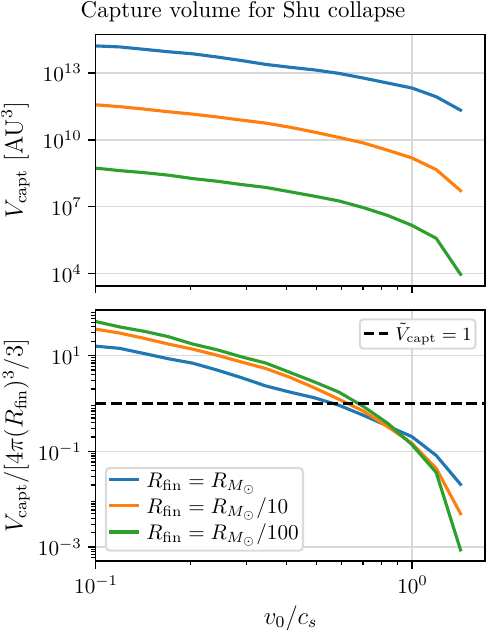}
    \caption{Top panel: volume $V_{\rm capt}$ of interstellar space from which planetesimals will be captured as a function of their initial velocity with respect to the cloud, normalized by the sound speed. This assumes that the potential evolves as in the Shu solution described in Section \ref{sect:Shu}. Different color curves correspond to different final expansion-wave radii $R_{\rm fin}=c_st_{\rm final}$; see text for details. Bottom panel: $V_{\rm capt}$ normalized by $4\pi R_{\rm fin}^3/3$. The horizontal dashed black line indicates a normalized capture volume of 1.}
    \label{fig:ShuCaptureRate}
\end{figure}

 
Figure \ref{fig:ShuCaptureRate} shows how the capture volume varies with $v_0$ for the Shu collapse.  In this case, the cutoff velocity is $1.4 c_s$ for all three clouds, roughly 25\% larger than the value predicted by Equation \eqref{eq:ShuVCrit}. The bottom panel shows the capture volume normalized by $4\pi R_{\rm fin}^3/3$, where $R_{\rm fin}=c_st_{\rm final}$.  As in the homogeneous and Larson-Penston cases, the capture volume is of order the cloud volume for $v_0 \approx c_s$.

\section{Largest body captured}
\label{sect:biggestCapture}

In this section we use the capture volumes calculated in Sections
\ref{sect:homoCollapse} and \ref{sect:sss} to estimate the largest planetesimal
that would typically be captured in each of our three collapse models.
To determine the number of captured bodies, we must assume a distribution
of initial speeds $v_0$ and a number density of planetesimals as a function
of size. We consider two cases: one in which the captured bodies come from the population of planetesimals freely floating about the Galaxy, and one in which they come from the birth cluster of the star.

\subsection{Average over a Maxwellian Distribution of $v_0$}
\label{sect:velavg}

As in Section \ref{sect:homoCollapse}, we assume that the initial speeds
$v_0$ of the planetesimals relative to the collapsing cloud follow a
Maxwellian distribution (see Equation \eqref{eq:Maxwellian}).  The mean speed of this distribution is given by $\langle v_0 \rangle = \sqrt{\frac{8}{\pi}} \sigma$. We then average the capture volume over this distribution to obtain

\begin{equation}
    \bar{V}_{\rm capt}(\sigma) = \int_{0}^{\infty}
    f(v_0;\sigma)\,V_{\rm capt}(v_0)\,\mathrm{d}v_0 .
    \label{eq:vavg}
\end{equation}

\begin{figure}[!htbp]
    \centering
\includegraphics[width=\columnwidth]{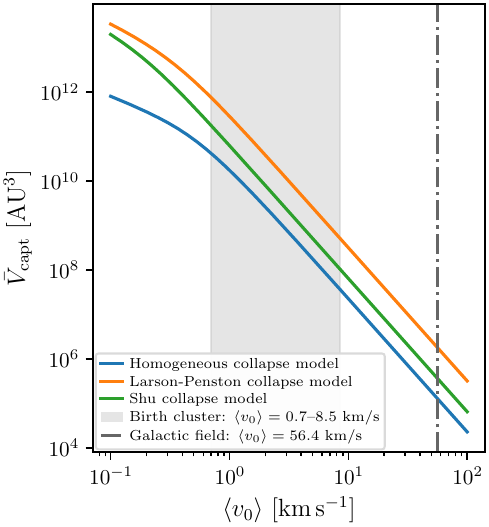}
    \caption{Maxwellian-averaged capture volume as a function of mean velocity for the three collapse models.  The shaded region corresponds to the range of velocity dispersions we consider in the birth-cluster environment, and the vertical dashed line corresponds to the Galactic value. }
    \label{fig:maxwellianCaptureVolume}
\end{figure}

Figure \ref{fig:maxwellianCaptureVolume} shows the capture volume as a function of $\langle v_0 \rangle$, assuming $v_0$ to have the distribution given in Equation \eqref{eq:Maxwellian}.  For each model, we took the parameters considered in Sections \ref{sect:homoCollapse} and \ref{sect:sss} that resulted in the largest capture volume; i.e. the blue curves in Figures \ref{fig:combinedCaptureVolume} - \ref{fig:ShuCaptureRate}.  The shaded region corresponds to the range of velocity dispersions we consider in the birth cluster environment, and the dashed line corresponds to the assumed value of $\langle v_0 \rangle$ in the Galaxy. 
\par
In general the speed at which a significant number of planetesimals are captured is small compared to $\langle v_0 \rangle$, and therefore $\bar V_{\rm capt}\propto\langle v_0 \rangle^{-3}$, since the exponential in the Maxwellian distribution approaches unity over the speeds contributing to capture.  This behavior is reflected in Figure \ref{fig:maxwellianCaptureVolume}.

\subsection{Size distribution}
\label{sect:sizedist}

We assume that the size distribution of interstellar planetesimal population has
the same shape as the size distribution measured for the long-period comet population by \citet{Boe19}. In terms
of the cometary nuclear absolute magnitude $H_N$, this is given by
\begin{equation}
    n(H_N) = n_1\, 10^{
    \int_{H_1}^{H_N}\alpha(H_N')\,\mathrm{d}H_N'} ,
    \label{eq:nH}
\end{equation}
where $n(H_N)$ is the number density per unit magnitude, and
$n_1$ is its value at the reference magnitude $H_1$. The slope $\alpha(H_N)$
is given by
\begin{equation}
    \alpha(H_N) = \alpha_{\rm small}
    + \frac{\alpha_{\rm big}-\alpha_{\rm small}}
    {1+\exp\left[\frac{H_N-H_{\rm trans}}{H_{\rm width}}\right]} ,
    \label{eq:alpha}
\end{equation}
so that the slope approaches $\alpha_{\rm big}$ for bright objects and
$\alpha_{\rm small}$ for faint objects. Following \citet{Boe19}, we assume
$\alpha_{\rm big}=0.70$, $\alpha_{\rm small}=0.07$, $H_{\rm trans}=16.9$, and
$H_{\rm width}=0.6$. For the assumed geometric albedo $p_V=0.04$, the
transition occurs at a diameter of approximately $2.8\,{\rm km}$.

We then convert this to a distribution in planetesimal diameter $D$ using
Equation 2 of \citet{Boe19},
\begin{equation}
    D(H_N,p_V) = \frac{1.342\times10^{6}\,{\rm m}}{\sqrt{p_V}}\,
    10^{-\frac{H_N}{5}} ,
    \label{eq:DofH}
\end{equation}
where $p_V$ is the geometric albedo. We can invert Equation \eqref{eq:DofH}
to obtain
\begin{equation}
    H_N(D) = 5\log_{10}\left(\frac{1.342\times10^{6}\,{\rm m}}
    {D\sqrt{p_V}}\right) .
    \label{eq:HofD}
\end{equation}
Using $\left|\frac{\mathrm{d}H_N}{\mathrm{d}D}\right|
= \frac{5}{D\ln 10}$, we can then write the differential number density
per unit diameter as
\begin{equation}
    \begin{aligned}
    n(D) & = n\big(H_N(D)\big)
    \left|\frac{\mathrm{d}H_N}{\mathrm{d}D}\right| \\
    & = \frac{5\,n_1}{D\ln 10}\,
    10^{\int_{H_1}^{H_N(D)}
    \alpha(H_N')\,\mathrm{d}H_N'} .
    \end{aligned}
    \label{eq:nD}
\end{equation}

We determine the normalization $n_1$ from the assumed mass density of
interstellar planetesimals $\rho_{\rm ptml}$. For spherical bodies, this
 is given by
\begin{equation}
    \rho_{\rm ptml} = \frac{\pi}{6}\rho_\bullet
    \int_0^{\infty} D^{3}\, n(D)\,\mathrm{d}D ,
    \label{eq:norm}
\end{equation}
where the bulk density of the planetesimals is
$\rho_\bullet = 3\,{\rm g\,cm^{-3}}$. The number density of planetesimals
with diameters larger than $D$ is then

\begin{equation}
	N(>D)=\int_D^\infty n(D')\,\mathrm{d}D'
\end{equation}

 Since the capture volume is independent of planetesimal size, we can multiply the Maxwellian-averaged capture volume given in Equation \eqref{eq:vavg} by $N(>D)$ to determine the expected number of captured planetesimals with diameters larger than $D$:

\begin{equation}
    \mathcal{N}(>D) = \bar{V}_{\rm capt}(\sigma)\, N(>D) .
    \label{eq:Ncapt}
\end{equation}
We define the characteristic largest captured diameter $D_{\rm largest}$ by
\begin{equation}
    \mathcal{N}(>D_{\rm largest}) = 1 .
\end{equation}

At large sizes, $n(D)\propto D^{-4.5}$, giving $\mathcal N(>D)\propto D^{-3.5}$, consistent with Figure \ref{fig:capturedSizeDistribution}. Thus $D_{\rm largest}\propto\bar V_{\rm capt}^{0.28}$.

\subsection{Maximum Size of Captured Planetesimals}
\label{sect:environments}

For the Galactic population, we assume that both the planetesimals and the newly-forming protostars have a Maxwellian velocity distribution with $\sigma_p$ = 25 km/s; roughly appropriate for the local stellar population \citep{Carlberg85,Yu18}. Assuming independent planetesimal and protostar velocities with equal dispersions gives a relative dispersion $\sigma=\sqrt{2}\sigma_{\rm p}
=35.4\ {\rm km\,s^{-1}}$ and mean relative speed $\langle v_0\rangle=\sqrt{8/\pi}\,\sigma=56.4\ {\rm km\,s^{-1}}$. Following \citet{Do18}, we normalize the size distribution to an interstellar planetesimal mass density of $\rho_{\rm ptml}=4\ M_\earth\,{\rm pc^{-3}}$.

For planetesimals from the birth cluster, we assume a lower mean speed and a larger number density. Characteristic 3D stellar mean speeds in open clusters are of order $0.5$--$6\ {\rm km\,s^{-1}}$ \citep{Angelo21}. Based on this, an an assumption that the ejection velocity of planetesimals from their host systems is negligible, the mean relative velocity should be a factor of $\sqrt{2}$ higher. We consider a slow model with $\langle v_0 \rangle = 0.7$ km/s, and a fast model with $\langle v_0 \rangle$ = 8.5 km/s. We assume a stellar number density of $n_\star=100\ {\rm pc^{-3}}$ \citep{Lada03} and assume that each star ejects $40\,M_\earth$ of planetesimals.  The typical age dispersion of a cluster is subject of debate \citep{Soderblom14}, and may be of order the disk lifetime. For this reason, we assume that at the time that a typical star is collapsing, only 20\% of the final free-floating planetesimal budget of the cluster is available for capture. These assumptions give $\rho_{\rm ptml}^{\rm clus}=8\times10^{2}\ M_\earth\,{\rm pc^{-3}}$, a factor of $200$ larger than the Galactic value.
\begin{figure*}[t]
    \centering
\includegraphics[width=\textwidth]{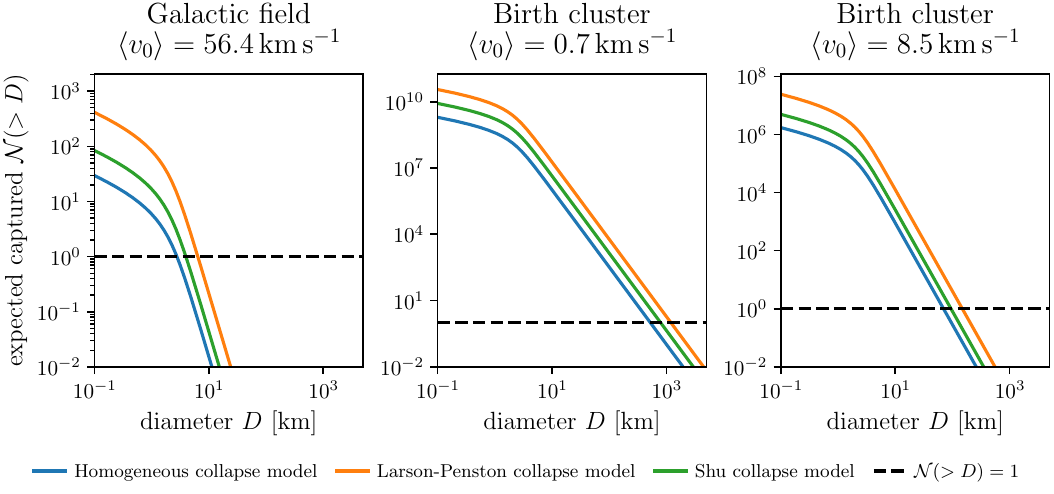}
    \caption{Expected number of captured planetesimals larger than diameter $D$ for homogeneous, Larson-Penston, and Shu collapse. From left to right, the panels show the Galactic field environment with $\langle v_0 \rangle = 56 \;{\rm km\,s^{-1}}$ and the two birth cluster models with $\langle v_0 \rangle = 0.7 \;{\rm km\,s^{-1}}$ and $\langle v_0 \rangle = 8.5\; {\rm km\,s^{-1}}$ The adopted planetesimal mass densities are $4\ M_\earth\,{\rm pc^{-3}}$ for the Galactic field and $800\ M_\earth\,{\rm pc^{-3}}$ for both birth-cluster panels.  The details of the collapse models are the same as in Figure \ref{fig:maxwellianCaptureVolume}. The horizontal dashed line marks one expected capture.}
    \label{fig:capturedSizeDistribution}
\end{figure*}
\par
 Figure \ref{fig:capturedSizeDistribution} shows $\mathcal{N} > D$ as a function of diameter for each of the three collapse models in each of our three model environments.  We find largest diameters of $2.8$--$6.4$ km for the Galactic population, $520$--$1200$ km for the lower-dispersion birth cluster, and $70$--$150$ km for the higher-dispersion cluster. Within each population, the characteristic largest captured diameters differ by approximately a factor of two between collapse models. 
\subsection{Implications for Planet Formation}
We find that the total mass budget of solid material within a protoplanetary disk is not significantly augmented by gravitationally captured planetesimals. Our model with the highest Maxwellian-integrated capture volume is the Larson-Penston collapse model with $\langle v_0 \rangle$ = 0.7 km/s.  In this case, the total captured mass is $M_{\rm capt}=\rho_{\rm ptml}^{\rm clus}\bar V_{\rm capt} =0.07\,M_\earth$. Since this is significantly less than the expected mass in a typical protoplanetary disk, and it is by no means certain that all of, or even a majority of the objects captured gravitationally will make it into the protoplanetary disk, we do not consider it plausible that gravitational capture of planetesimals significantly impacts the overall solid abundance in the disk.
\par
On the other hand, a single sufficiently large planetesimal, if entrained within the flow of the local disk material, may grow to planet size within the disk lifetime by accretion of pebbles that formed within the disk.  \citet{Lambrechts12} and \citet{Lyra23} study the rate at which planetesimals may grow as a result of collisions with small aerodynamically coupled solid objects, in the process known as ``pebble accretion".  They find that growth to planet size within 1 million years is possible at 5 AU from the central star starting from a seed planetesimal with diameter between a few hundred and about a 1000 km, depending on the assumed disk mass, pebble size distribution, and disk pressure profile.  These numbers are comparable to or smaller than than what we reported in the previous section for the largest body captured in the case in which the planetesimals come from the birth cluster, but much larger than the largest body reasonably captured from the Galactic planetesimal population.  Thus, we conclude that gravitational capture of planetesimals from the same birth cluster may accelerate planet formation in some stars, or even enable it to occur even in disks that are, for whatever reason, not efficient at forming their own planetesimals. 
\par
The 1D collapse models studied here are not sufficient to determine whether the captured planetesimals actually end up entrained in the disk material.  Thus, our estimate of the largest body captured should be taken as an upper bound on the largest body that is entrained in the disk material.  This question, as well as the possibility of erosion of the planetesimals by sand-blasting with pebbles as they interact with the disk will be investigated in the future.

\section{Conclusion}
\label{sect:conclusion}
We calculate the rate at which freely floating planetesimals are captured into orbit around a newly-formed star due
to the time-varying potential of the collapsing prestellar core.  We considered three different collapse models: collapse of a homogeneous spherical density distribution at a fixed fraction of the free-fall rate, as well as the self-similar Larson-Penston and Shu 1D collapse models.  We pay special attention to the size of the largest captured body, because a single planetesimal, if sufficiently large, could grow to planet size via pebble accretion.  The total mass of solid material captured is never more than a small perturbation on the solid mass present in a disk, and not enough to grow a planet.  We find the following results:
\begin{enumerate}
    \item The capture rate depends sharply on the initial velocity.  In the homogeneous collapse model, the capture rate declines as $v_0^{-6}$ for values of $v_0$ large compared to the escape speed of the cloud at the beginning of the modeled collapse. The Larson-Penston and Shu models predict that the capture rate drops to zero for initial speeds greater than 5.9 (Larson-Penston), or 1.4 (Shu) times the sound speed.
    \item As a result of the strong dependence of the capture rate on the initial velocity, the number of captured planetesimals is highly sensitive to the assumed velocity dispersion.
    \item Our modeling suggests that it is very unlikely that a forming star will capture an object from the Galactic planetesimal population large enough to impact the planet formation process.  Depending on the collapse model assumed, we find a maximum size captured between 2.8 and 6.4 km.  
    \item We consider also the possibility that planetesimals could be captured from the birth cluster of the forming star.  This mechanism appears much more promising, with a maximum capture size varying between 70 and 1,200 km, depending on the assumed collapse model, and assuming mean relative speeds between 0.7 and 8.5 km/s.
    \item The total mass of planetesimals that can be gravitationally captured is not enough to significantly alter the planet formation process unless a single large body is captured.  Even under the most favorable scenario that we considered, only 0.07 Earth masses of solid material is captured.  
    \item We emphasize that our results are only estimates for the distribution of objects that are captured into orbit around the forming protostar.  Further work is needed to determine what fraction of the planetesimals captured in orbit around the star would actually become entrained in the disk material. Orbital evolution via gas drag and erosion via collisions with pebbles present in the host disk, both processes not considered in this work, would likely play a significant role in this process.
\end{enumerate}

{\hspace{-.45 cm} \bf Acknowledgments:}
\\
KS gratefully acknowledges helpful discussions with Jonathan Squire, Wladimir Lyra and Susanne Pfalzner.

\bibliographystyle{aasjournalv7.1}
\bibliography{report_new}

\end{document}